# Resource Management optimally in Non-Orthogonal Multiple Access Networks for fifth-generation by using game-theoretic

Mahsa Khodakhah[1], Nahid Ardalani[2]

[1,2] Islamic Azad University, Central Tehran Branch, Tehran, Iran

E-mail:mahsa.khodakhah@gmail.com[1]

N_ardalani@yahoo.com[2]

**Abstract:**

In this paper , we optimize the resource Allocation management by using game-Theory . In this paper we can optimize number of users in resource allocation management by using the user-sub-channel-soap matching algorithm (USMA) for non-orthogonal multiple access for fifth-generation networks by increase multiple access user and NOMA that is to accessible to 63 users. This method reduces interference between users , which include costs , and resource to access for other users.

## I. Introduction:

Non-orthogonal multiple access techniques have been proposed recently for 5G wireless systems and beyond . NOMA is an in-cell multi-user sharing scheme that has not been used in the field of power and code in the previous generation .In [1] the domain of code, time and frequency was a characteristic, than NOMA can supports simultaneously a large number of users. NOMA allows multiple users to assign a sub-carrier to multiple users, creating multi-user interference. For this problem, the Multi-User Detection (MUD) technique, such as SIC interception on the receiving side, is used to decode incoming signals. NOMA utilizes the intelligent reuse of network resources and the integration of multi-user messages in the same transmitter's sub-channel. Resource management is divided into two categories: the power domain and the code domain .Now, using game theory, we were able to improve the total sum-rate and improve the performance of the system[1]-[2]. In this paper, using the USMA algorithm, we were able to optimize resource allocation. The rest of this article is below . In section II we describe model system .In section III, we describe MTM Matching Algorithm for NOMA , simulation results are presented in section IV .Finally , we conclude the paper in section V.

## II. Model system:

In [2,4-6] sparse code multiple access (SCMA) in a single-cell uplink network is shown in fig1 . Transfer of a set of mobile users N={1,2,…,N}to a single BS . The exiting bandwidth is divided into a set of subcarrier, which is represented as K={1,2,…,k}.

$$max_{f_{kj},p_{kj}} \sum_{j \in N} \sum_{k \in K} f_{k,j} \log_2(1 + \frac{p_{kj}|h_{kj}|^2}{\sigma_n^2 + I_{kj}})$$

Figure1: NOMA network system model

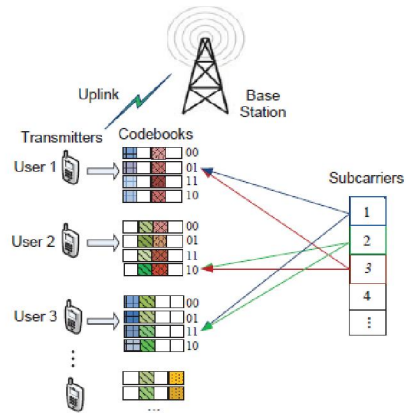

### III . MTM Matching Algorithm for NOMA

Matches can be divided into 3 parts: one-to-one matching , one-to-many matching , many-to-many(MTM) matching. Each user matched a user sub-carrier. First, each player's matching list is introduced . This matching user wraps with another matching user and creates a pair of block swaps. This swap-blocking is exchanged with the matching of those matching users. This will happen as long as the current implementation is being updated .This process is determined at this stage of the finalization until it is repeated that no user can find any new swap-blocking pair. The USMA algorithm(Table1) is presented for solving the resource allocation problem in Part II. In the first stage, the twenty matches are created for each player to apply for their priority and equal power allocation by each user. We use USMA algorithm and an initial adaptive structure is formed by a random matching adaptation of $d_v$ and $d_f$ . Including multiple replicas that each user submits to the search for another user creates a pair of swap-blocking then they exchange their matching and update the current matching . A repetition stops when no user can find the pair of new swap-blocking and a final match is determined. And in the end, each user performs the power allocation function [2]-[3].

TABLE I
USER-SUBCHANNEL SWAP-MATCHING ALGORITHM (USMA)

**Step 1: Startup.**
(a) Construct the matching list for each user $N_j$ and subcarrier $SC_k$, denoted by $\mathcal{L}_{N_j}$ and $\mathcal{L}_{SC_k}$. Set $\mathcal{L}_{N_j} = \mathcal{K}_{SC}$, and $\mathcal{L}_{SC_k} = \mathcal{N}$.
(b) The transmitted power of each user is equally allocated to its matched subcarriers.
(c) Construct an empty matching $\Psi$. Users and sub-channels are randomly matched with each other subject to $|\Psi(SC_k)| \leq d_f$ and $|\Psi(N_j)| \leq d_v$.

**Step 2: Swap matching phase.**
For each matched user $N_j$ in $\Psi$, it keeps searching $\mathcal{N}\setminus\{N_j\}$ for a swap-blocking pair $(N_i, N_j)$ along with $SC_p \in \Psi(N_i)$ and $SC_q \in \Psi(N_j)$.
(i) If there exists such a swap-blocking pair, $N_i$ exchanges its match $SC_p$ with $N_j$ for $SC_q$. Set $\Psi = \Psi_{jq}^{ip}$.
(ii) Else, $N_j$ keeps its matches.
*until* no user can form a swap-blocking pair with any other users.

**Step 3:** *End of the algorithm.*

## IV. Simulation Results

The problem can be optimized to get f Matrix to used USMA algorithm. In f matrix the number mobile is players problem. In each row, the codes given to users are maxed out to 5 code 1, and the rest of the layers are zero, and in each column, a maximum of 3 code 1 and the remainder of the layers is zero in the sub-carrier. This matrix is randomly designed, that is, the matrix of each series that runs the simulation gives a matrix randomly. Users programmed have been checked against the number of users in a time slot .NOMA design, considering the complexity of decoding and signal costs for base station receivers, is considered as 10 sub-channels . We assume that the path is reduced by using the modified HATA model and the users are distributed uniformly in a square area of 350 meters in length. Each sub-channel can only be assigned a maximum user $D_F$ and each user can only occupy a maximum of five sub channels. Using the algorithm subcarrier-user USMA matrix $F = [f_{ij}]$ that use optimization is achieved. In the conditions of the problem, we assume in the problem that the drop in the electromagnetic wave power density released in space is obtained from the HATA urban model, and the users are distributed uniformly. HATA's model is a radio broadcast to predict the loss of the cellular transfer path in outer environments, which is valid for microwave frequencies of 150 to 1500 $MHz$. This is an empirical formula based on data Okumura model is so commonly referred to as the HATA-Okumura . The graphic information on behalf of the Okumura model and the more it will jump to realize the effects of reflection and scattering of structures makes city. In addition, in [1] HATA reform models for application programs in suburban and rural environments is applied. By comparing the simulation to simulate the proposed reference to the results that we've achieved over the reference optimal solutions. As a result, users access to sub-carriers is better because of the accidental results of the matrix f. Figure 2 shows that the number of users scheduled increases with an increase in the number of users and is much higher than the number of subscribers. Figure 3 determine the number of users scheduled as maximum, minimum and average cases.

Table 2: $d_v = 3, d_f = 5$

| N | 10 | 20 | 30 | 40 | 50 | 60 | 70 | 80 | 90 | 100 |
|---|---|---|---|---|---|---|---|---|---|---|
| Refrense1 | 10 | 18 | 23 | 26 | 29 | 30 | 31 | 32 | 33 | 34 |
| Proposed reference | 6.58 | 14.61 | 19.18 | 21.76 | 28.50 | 41.90 | 41.96 | 43.62 | 58.31 | 63.25 |

Figure 2: Suggested simulation with d_f=5 ,d_v=3

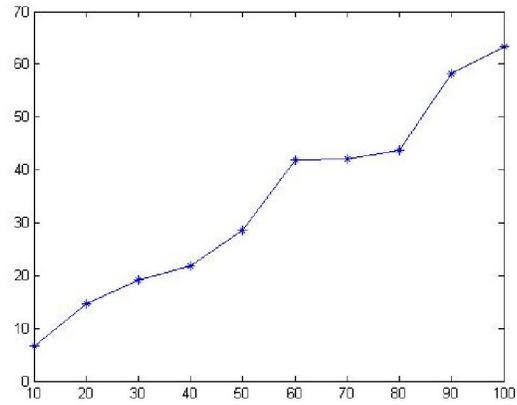

Figure 3

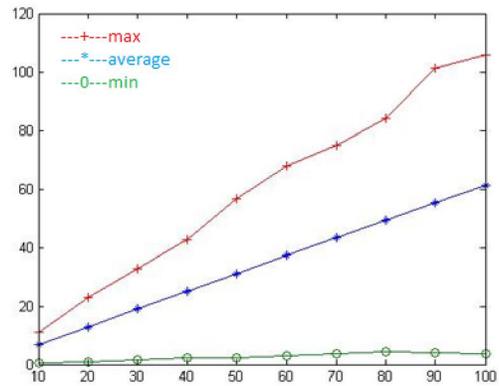

**Figure 4**

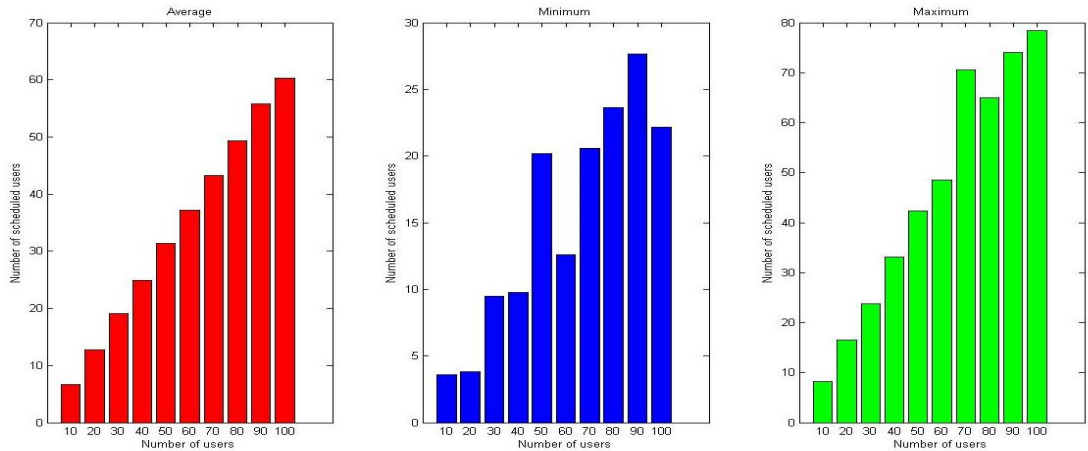

## V. Conclusion and Future work

In this thesis, using the domain-domain and the adaptive game that is considered as a set of users and sub-integrals, in order to maximize the total profit rate by using the user-to-subtraction-matching algorithm, The sub-channel, in which the number of users in resource allocation management has been optimized with non-orthogonal access method for fifth-generation networks .For example, in 100 users, the number of optimized users has reached 63 users in terms of maximum profits .Given that the matrix f presented in the previous section is random, the number of possible modes for the matrix f is equal to (100| 5) (8/3).The average results obtained is the optimal answer to the problem presented by our proposed solution .Our proposed solution is better than the original reference, our proposed approach to the OFDM scheme is also better .Using the USMA algorithm and the game theory , we were able to improve the original reference result shown in the proposed simulation.

# REFRENCE